\documentclass[aps,preprint,floatfix,nofootinbib]{revtex4}
\usepackage{epsf}
\usepackage{graphicx}

\def\lmu{\left|U^L_{\mu N}\right|^2}
\def\rmu{\left|U^R_{\mu N}\right|^2}
\def\lt{\left|U^L_{\tau N}\right|^2}
\def\rt{\left|U^R_{\tau N}\right|^2}
\def\le{\left|U^L_{e N}\right|^2}
\def\re{\left|U^R_{e N}\right|^2}
\def\rmus{\left|U^R_{\mu N}\right|^4}
\def\res{\left|U^R_{e N}\right|^4}

\def\ln#1{U^{L*}_{#1 N}}
\def\rn#1{U^R_{#1 N}}
\def\lns#1{U^{L}_{#1 N}}
\def\rns#1{U^{R*}_{#1 N}}

\def\lni#1#2{U^{L*}_{#1 N_#2}}
\def\rni#1#2{U^{R}_{#1 N_#2}}
\def\lnis#1#2{U^{L}_{#1 N_#2}}
\def\rnis#1#2{U^{R*}_{#1 N_#2}}

\begin{document}

\title{Lepton flavor violating $\tau$ and $B$ decays and heavy neutrinos}

\author{Xiao-Gang He$^{a,b}$,\ G. Valencia$^c$,\ Yili Wang$^c$}

\email{hexg@phys.ntu.edu.tw}
\email[]{valencia@iastate.edu}
\email[]{yiliwa@iastate.edu}

\affiliation{$^a$
Department of Physics, Peking University, Beijing, China\\
$^b$Department of Physics, National Taiwan University, Taipei\\
$^{c}$Department of Physics, Iowa State University, Ames, IA 50011}

\date{\today}

\vskip 1cm
\begin{abstract}

We study lepton flavor violating (LFV) $\tau$ and $B$ decays in models with heavy neutrinos to constrain the mixing matrix parameters $U_{\tau N}$. We find that the best current constraints when the heavy neutrinos are purely  left-handed  come from LFV radiative $\tau$ decay modes. To obtain competitive constraints in LFV $B$ decay it is necessary to probe $b\to X_{s} \tau^{\pm} e^{\mp}$ at the $10^{-7}$ level. When the heavy neutrinos have both 
left and right-handed couplings, the mixing parameters can be constrained by studying LFV $B$ decay modes and LFV $\tau$ decay into three charged leptons.  We find that the branching ratios $B(\tau^{\pm}\to\ell_{1}^{\pm}\ell_{2}^{\pm}\ell_{3}^{\mp})$, $B(B_{s}\to \tau^{\pm} e^{\mp})$ and 
$B(b\to X_{s} \ell_{1}^{\pm}\ell_{2}^{\mp})$ need to be probed at the $10^{-8}$ level in order to constrain the mixing parameters beyond what is known from unitarity.

\end{abstract}

\pacs{PACS numbers: }

\maketitle

\section{Introduction}

In this paper we study lepton flavor violating $\tau$ and $B$
decays. In the minimal Standard Model(SM), generation lepton
number is conserved. However, the observation of neutrino oscillations
implies that family lepton number must be violated \cite{neutrino}. At present it is not clear if the total lepton number is violated. The neutrino
oscillation is due to a mismatch between the weak and mass
eigenstates of neutrinos. This mismatch causes mixing between
different generations of leptons in the charged current interaction
with the $W$ boson. In principle, flavor changing neutral current  (FCNC) processes 
in the lepton sector occur as well. Some examples would be 
$\tau \to \ell \gamma$, 
$\tau \to \ell_1\ell_2\bar \ell_3$, $B\to \ell \bar \ell^{\prime}$ 
and $B \to \ell \bar \ell^{\prime} X_s$.
Although no direct experimental evidence for such FCNC exists, there are experimental constraints \cite{Abe:2003sx,Edwards:1996te,Bliss:1997iq,Aubert:2003pc,Yusa:2004gm,Aubert:2004gm,Chang:2003yy,Bornheim:2004rx,
Glenn:1997gh,Aubert:2002aw}. 
The decays $\tau \to \mu (e) \gamma$ have recently been the subject 
of considerable attention \cite{recentw}. 
They have been studied in connection with
LFV occurring through mixing with heavy neutrinos in the context of 
supersymmetric theories where these modes are found to be a promising 
tool to constrain the models.
In this paper we investigate the potential rates for these processes 
in Left-Right (LR) models with heavy neutrinos. 

FCNC in the lepton sector that are solely 
due to mixing in the charged current interaction with  the
usual left-handed W boson and light neutrinos are extremely small because they are suppressed by powers of $m^2_\nu/M^2_W$ \cite{Inami:1980fz} . 
One way to increase the FCNC 
interaction in the lepton sector is to introduce heavy neutrinos so 
that the suppression factor $m^2_\nu/M^2_W$ is not in effect. This
can be done, for example, by introducing a heavy fourth generation. 
If one insists on having just three light left-handed neutrinos, one
needs to give the right-handed neutrinos heavy Majorana masses.
The heavy neutrino can appear through mixing in the charged
current interaction and enhance the FCNC interaction in the lepton sector. 
The introduction of right-handed neutrinos also raises the possibility
of having right-handed charged currents by adding to the theory
a right-handed $W^\prime$ boson. This new charged
current interaction can lead to additional effects in the above
decay processes and here we consider such a possibility  \cite{pilaftsis}.

A natural model of this type is the LR model
based on the $SU(3)_c\times SU(2)_L\times SU(2)_R\times U(1)_{B-L}$
gauge group \cite{lrmodel}. In the Left-Right model, FCNC interactions arise from several sources. In this paper we consider the  exchange 
of $W_{L,R}$ bosons at one loop level. It is well known that to obtain 
gauge invariant results one must also include charged Higgs boson 
effects at one loop level as well as 
tree level exchange of neutral Higgs bosons \cite{probslr}. We will comment on these effects but will not discuss them in detail as they depend on several unknown parameters. We concentrate on the effects that 
depend only on the $W^{\prime}$ mass and ignore those that depend on Higgs boson masses to illustrate the constraints that can be placed on the mixing with heavy Majorana 
neutrinos by the processes $\tau \to l \gamma$, 
$\tau \to \ell_1\ell_2\ell_3$, $B\to \ell \bar \ell^{\prime}$, 
or $B \to \ell \bar \ell^{\prime} X_s$. Our paper complements existing studies of the modes $\mu\to e\gamma$ and 
$K_{L}\to \mu e$ \cite{pilaftsis}, and extends them to include the mixing parameters 
$U_{\tau N}$. 

In LR models there are left-handed light neutrinos $\nu_L$
and right-handed heavy neutrinos $\nu^{\prime}_R$, and these neutrinos can
be Majorana particles. If there are three light and N heavy
neutrinos, the general mass term for the neutrinos can be written
as
\begin{eqnarray}
L_M = -{1\over 2} (\bar \nu_L, \bar \nu^{'c}_R) M^\nu \left (
\begin{array} {l}
\nu^c_L\\
\nu^{'}_R
\end{array}
\right ) + H.C.
\end{eqnarray}
$M^\nu$ is a symmetric matrix which can be diagonalized
\begin{eqnarray}
\tilde U^T M^\nu \tilde U = \hat M^\nu,
\end{eqnarray}
with the aid of a unitary matrix $\tilde U$ resulting in $\hat M^\nu = diag(m_1,m_2,m_3, M_4, M_5, \cdots)$ with $m_i$ and $M_i$ denoting the light and heavy mass eigenvalues respectively. 

If there is no right-handed W-boson interaction, the number
of right-handed heavy neutrinos is unrelated to the number of 
charged leptons. However, when a right-handed W-boson is
introduced and the heavy neutrinos are required to interact with it, 
it is natural to have the heavy neutrinos and the right-handed
charged leptons form $SU(2)_R$ doublets. 
In this case there are as many heavy neutrinos as charged 
leptons (three).

The most general charged current interactions of charged leptons
and neutrinos with W-bosons can be parameterized in the weak
interaction basis, as
\begin{eqnarray}
{\cal L}_{lepton} &=& -{g_L\over \sqrt{2}}W^\mu \bar L \gamma_\mu
\left(g^\ell_LP_L +
g^\ell_R P_R \right )\nu\nonumber\\
&-&{g_R\over \sqrt{2}}W^{\prime\mu} \bar L \gamma_\mu \left(\tilde
g^\ell_LP_L + \tilde g^\ell_R P_R\right)\nu^{\prime};\nonumber\\
{\cal L}_{quark} &=& -{g_L\over \sqrt{2}}W^\mu \bar U \gamma_\mu
\left(g^d_LP_L +
g^d_R P_R\right)D\nonumber\\
&-&{g_R\over \sqrt{2}}W^{\prime\mu} \bar U \gamma_\mu \left(\tilde
g^d_LP_L + \tilde g^d_R P_R\right) D,
\end{eqnarray}
where $L = (e,\;\;\mu,\;\;\tau)^T$, $\nu =
(\nu_e,\;\;\nu_\mu,\;\;\nu_\tau)^T$, $\nu' =
(\nu'_e,\;\;\nu'_\mu,\;\;\nu'_\tau)^T$, $U = (u,\;\;c,\;\;t)^T$,
and $D = (d,\;\;s,\;\;b)^T$. In the above $W$ and $W^\prime$ denote 
the mass eigenstates of W-bosons with $W$ being mostly left-handed 
and $W^\prime$ being mostly right-handed.

In the mass eigenstate basis, we have  \cite{pilaftsis}
\begin{eqnarray}
{\cal L}_{lepton} &=& -{g_L\over \sqrt{2}}W^\mu\sum_{l=1}^3\bar
\ell^m_l
\gamma_\mu \left(g^\ell_LP_LU^{L*}_{\ell j} + g^\ell_R P_RU^R_{\ell j}\right)\nu^m_j\nonumber\\
&-&{g_R\over \sqrt{2}}W^{\prime\mu} \sum_{l=1}^3\bar \ell^m_{l}
\gamma_\mu \left(\tilde g^\ell_LP_L U^{L*}_{\ell j} + \tilde g^\ell_R
P_R U^R_{\ell j}\right)\nu^m_j; \nonumber\\
{\cal L}_{quark} &=& -{g_L\over \sqrt{2}}W^\mu \bar
U^m \gamma_\mu \left(g^d_LP_L V^L +
g^d_R P_R V^R \right)D^m\nonumber\\
&-&{g_R\over \sqrt{2}}W^{\prime\mu} \bar U^m \gamma_\mu
\left(\tilde g^d_LP_LV^L + \tilde g^d_R P_RV_R\right)D^m.
\end{eqnarray}
The left-handed and right-handed charged leptons are diagonalized by the matrices $S^{L,R}$: $\ell^m_L = S^L
\ell_L$ and $\ell^m_R = S^R \ell_R$ and 
we have defined the matrices  $U^{L*}_{\ell j} = \sum_{i=1}^3
S^{\dagger L}_{\ell i}\tilde U^*_{ij}$ and $U^R_{\ell j} = \sum_{l'=1}^3 S^{\dagger R}_{\ell \ell^\prime}\tilde U_{(\ell^\prime + 3)j}$ with  $\ell = e$, $\mu$ and $\tau$ . 

In what follows we will  drop the superscript ``m''
from the fermion fields and always refer to mass eigenstates. Note that $U^{L,R}$ are $3\times 6$ matrices and we construct the matrix 
\begin{eqnarray}
U^\prime = \left ( \begin{array}{l} U^{L}\\U^R
\end{array}
\right ),
\end{eqnarray}
with
\begin{eqnarray}
U^L & =& \left ( \begin{array}{llllll}
U^L_{e 1}&U^L_{e 2}&U^L_{e 3}&U^L_{e 4}&U^L_{e 5}&U^L_{e 6}\\
U^L_{\mu 1}&U^L_{\mu 2}&U^L_{\mu 3}&U^L_{\mu 4}&U^L_{\mu 5}&U^L_{\mu 6}\\
U^L_{\tau 1}&U^L_{\tau 2}&U^L_{\tau 3}&U^L_{\tau 4}&U^L_{\tau 5}&U^L_{\tau 6}\end{array}
\right ) {\rm ~~and} \nonumber\\
U^R & =& \left ( \begin{array}{llllll}
U^R_{e 1}&U^R_{e 2}&U^R_{e 3}&U^R_{e 4}&U^R_{e 5}&U^R_{e 6}\\
U^R_{\mu 1}&U^R_{\mu 2}&U^R_{\mu 3}&U^R_{\mu 4}&U^R_{\mu 5}&U^R_{\mu 6}\\
U^R_{\tau 1}&U^R_{\tau 2}&U^R_{\tau 3}&U^R_{\tau 4}&U^R_{\tau 5}&U^R_{\tau 6}\end{array}
\right ).
\end{eqnarray}
This can be viewed as the unitary matrix which diagonalizes the neutrino
mass matrix in the basis where the charged lepton mass matrix is
already diagonal. The following relations hold
\begin{eqnarray}
&&\sum_{j=1}^{6} U^{L*}_{\ell j} U^L_{\ell^\prime j}
=\delta_{\ell\ell^{\prime}} ,\;\;\;\; \sum_{j=1}^{6} U^{R*}_{\ell
j} U^R_{\ell^\prime j} = \delta_{\ell\ell^{\prime}} ,\nonumber\\
&&\sum_{j=1}^{6} U^{L}_{\ell j} U^{R*}_{\ell^\prime j} =0,\;\;\;\;
\sum_{\ell= e,\mu,\tau} U^{L*}_{\ell j} U^L_{\ell i} + \sum_{\ell =e,\mu,\tau}
U^{R*}_{\ell i}U^R_{\ell j} =\delta_{ij}.
 \label{unitarity}
\end{eqnarray}

There is some information on the matrix elements $U^L_{e2}$ and $U^L_{\mu3}$ from neutrino oscillation experiments \cite{neutrino}, which prefer them to be in the ranges $0.50 \sim 0.69$ and $0.60\sim 0.80$ respectively. 
For the processes that we discuss in this paper there can only be large effects if there is substantial mixing with the heavy neutrinos and there are few constraints on these parameters. In see-saw models, the generic size of the matrix 
elements $U^L_{\ell, (4,5,6)}$ and 
$U^R_{\ell, (1,2,3)}$ is of order $m_D/M_N$
where $m_D$ is the Dirac neutrino mass and $M_N$ is
the heavy Majorana neutrino mass. In such a scenario the
light neutrino masses are typically $m^2_D/M_N$. This requires 
the heavy neutrino masses to be heavier than a few hundred
GeV, and results in the above mixing matrix elements being extremely 
small. In these models, contributions from one $W$ exchange to 
radiatively  induced penguin processes, and from box diagrams 
with two $W$'s or one $W$ and one $W'$ exchanges are too small to 
be observed. Even in these models, however, the 
matrix elements $U^R_{l, (4,5,6)}$ can be of 
order one. For this type of model the exchange of $W^\prime$'s 
in both radiative penguin and box processes may produce 
observable effects. There are also special cases in which the elements
$U^L_{l,(4,5,6)}$ and $U^R_{l,(1,2,3)}$ can be sizeable. An
example has been discussed in Ref.~\cite{pilaftsis}, 
where some of these matrix elements are of order
$m_D/M_N$, but the light neutrino masses, at tree level, 
are not directly related to them and the ratio $m_D/M_N$ does not 
need to be very small. Since our aim is not to study specific models,
but to provide an estimate of the sensitivity needed in LFV $\tau$ 
and $B$ decay modes in order to constrain the general Left-Right
model mixing beyond the requirement of unitarity, we will treat 
the matrix elements in $U^\prime$ as arbitrary in this paper.

The couplings $g^{\ell,d}_{L,R}$ and $\tilde g^{\ell,d}_{L,R}$ are in
general complex numbers. In renormalizable models without
left-right gauge boson mixing, $g^{\ell,d}_L = 1$, $g^{\ell,d}_R = 0$,
$\tilde g_L^{\ell,q} = 0$ and $\tilde g_R^{\ell,q} =1$. If there is
left-right gauge boson mixing with a mixing angle $\xi_W$  ($W=W_L \cos\xi_W +W_R
\sin\xi_W$ and $W' = -W_L \sin\xi_W + W_R \cos\xi_W$) then 
\begin{eqnarray}
&&g^\ell_L = g^d_L = \cos\xi_W,\;\;g^\ell_R = g^d_R = {g_R\over
g_L}\sin\xi_W;\nonumber\\
&&\tilde g^\ell_L = \tilde g^d_L = -{g_L\over g_R}\sin\xi_W,\;\;\tilde
g^\ell_R = \tilde g^d_R = \cos\xi_W.
\label{mixcoups}
\end{eqnarray}
Throughout the paper we will use the notation: 
\begin{equation}
\lambda_i \equiv {m_i^2 \over M_{W}^2},\, \ \
\beta\equiv {M^2_W \over M^{\prime 2}_W},\, \ \
\xi_g = {g_R \xi_W \over g_L},\, \ \
\beta_g \equiv  {g_R^2 \beta \over g_L^2}.
\label{notation}
\end{equation}

For our loop calculations we will assume that all 
fermions are massless except
for the top-quark and the heavy right-handed neutrinos. We will also assume that $\beta$ is smaller than a few percent in keeping with bounds on $W^{\prime}$ bosons \cite{Eidelman:2004wy}. Finally, we will assume that $W_{L}-W_{R}$ mixing is small as indicated by $b \rightarrow s \gamma$. In particular, following  \cite{Chen:2001fj} we found
in  \cite{He:2002ha} at the $2\sigma$ level that there are
two allowed ranges for $\xi_g$. They correspond to destructive and
constructive interference with the standard model amplitude respectively and are 
\begin{eqnarray}
-0.032 < & {V^R_{tb}\over V^L_{tb}}\xi_g & < -0.027, \nonumber \\
-0.0016 < & {V^R_{tb}\over V^L_{tb}} \xi_g & < 0.0037.
\label{ranxief}
\end{eqnarray}
Overall, $\xi_g$ is constrained to be very small and in our analysis 
we will only keep terms linear in $\xi_g$ in the matrix elements. 

We will use this framework to examine LFV  induced by neutrino mixing (with new heavy neutrinos). Our purpose is to provide an estimate of the sensitivity needed in LFV $\tau$ and $B$ decay modes in order to constrain this scenario beyond the requirement of unitarity of the matrix $U^{\prime}$.   
 
\section{LFV radiative $\tau$ decay}

\begin{figure}[!htb]
\begin{center}
\includegraphics[width=10cm]{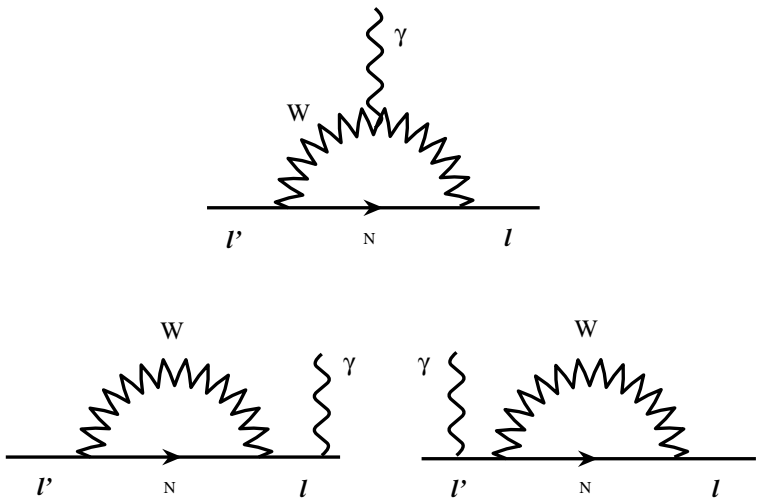}
\end{center}
\caption{Diagrams giving rise to $\ell^\prime\rightarrow \ell \gamma$ 
in unitary gauge.}
\label{f:lplg}
\end{figure}

Beginning with $\mu \rightarrow e \gamma$  \cite{mueg}, processes 
of the form $\ell^\prime\rightarrow \ell \gamma$ have been used to 
constrain new physics, including heavy neutrinos. Here we consider the case of $\tau$ decay. The one-loop effective operator can be 
calculated in unitary gauge from the diagrams in Figure~\ref{f:lplg}. 
We first consider the case of a very heavy $W^{\prime}$ so that only the $W$ is exchanged in the loop. In keeping with current 
experimental constraints Eq.~\ref{ranxief}, we work only to first 
order in the  $W_L-W_R$ mixing parameter $\xi_W$.
This  results in two operators which we write in the form
\begin{eqnarray}
{\cal L} &=& 4 {G_F \over \sqrt{2}} {e \over 16\pi^2} F^{\mu\nu}
\sum_N \left[
 \ln{\ell} \lns{\ell^\prime} F(\lambda_N)\
m_{\ell^\prime} \bar\ell \sigma_{\mu\nu} P_R \ell^\prime \right . \nonumber \\
& + & \left .
{g_R \over g_L} \xi_W \tilde{F}(\lambda_N) M_N \bar\ell \sigma_{\mu\nu} \left( \rn{\ell}\lns{\ell^\prime} P_L + \ln{\ell}\rns{\ell^\prime} P_R \right)\ell^\prime \right].
\label{lplgops}
\end{eqnarray}
Here $F^{\mu\nu}$ is the electromagnetic field strength tensor and
the Inami-Lim functions are given by
\begin{eqnarray}
F(\lambda_N) &=& \left[ {3 \lambda_N^3\log\lambda_N \over
4(1-\lambda_N)^4} + {2\lambda_N^3+5\lambda_N^2-\lambda_N
\over 8(1-\lambda_N)^3}\right],
\nonumber \\
\tilde{F}(\lambda_N)&=& \left[ {3 \lambda_N^2\log\lambda_N
\over 2(1-\lambda_N)^3} - {\lambda_N^2-11\lambda_N +4\over
4(1-\lambda_N)^2}\right].
\nonumber \\
\label{lplgcoef}
\end{eqnarray}
The exchange of a $W^\prime$ can be easily included and it 
leads to similar expressions:
\begin{eqnarray}
{\cal L} &=& 4 \beta {G_F \over \sqrt{2}} {e \over 16\pi^2} F^{\mu\nu}
\sum_N \left[
 \rn{\ell} \rns{\ell^\prime} F(\beta\lambda_N)\
m_{\ell^\prime} \bar\ell \sigma_{\mu\nu} P_L \ell^\prime \right.\nonumber \\
& - & \left . {g_L \over g_R} \xi_W \tilde{F}(\beta\lambda_N) M_N \bar\ell \sigma_{\mu\nu}\left( \rn{\ell}\lns{\ell^\prime} P_L +  \ln{\ell}\rns{\ell^\prime} P_R \right)\ell^\prime \right].
\label{lplgopsr}
\end{eqnarray}

We now calculate the branching ratio for $\ell^\prime\rightarrow\ell\gamma$ neglecting $m_\ell$. 
We first consider the case without $W_L-W_R$ mixing, dominated by the first operator in Eq.~\ref{lplgops}. We find
\begin{equation}
\Gamma(\ell^\prime \rightarrow \ell \gamma) = {G_F^2 \alpha \over
32 \pi^4} m_{\ell^{\prime}}^5 \left| \sum_N \ln{\ell} \lns{\ell^\prime}
F(\lambda_N)\right|^2.
\label{lptolg}
\end{equation}
For $\tau$ decay, it is convenient to express this result as a fraction of the rate \begin{equation}
\Gamma(\tau^- \rightarrow \mu^- \nu_\tau \bar\nu_\mu) = {G_F^2 m_\tau^5 \over 192 \pi^3},
\label{gamt}
\end{equation}
to obtain for $\ell = \mu,\ e$,
\begin{equation}
R_\ell \equiv {\Gamma (\tau^- \rightarrow \ell \gamma) \over
\Gamma(\tau^- \rightarrow \mu^- \nu_\tau \bar\nu_\mu)}
= \left({6\alpha \over \pi}\right)\left| \sum_N \ln{\ell} \lns{\tau}
F(\lambda_N)\right|^2.
\end{equation}

The current experimental bounds
$B(\tau \rightarrow \mu \gamma) \leq 3.1 \times 10 ^{-7}$ 
 \cite{Abe:2003sx},
$B(\tau \rightarrow e \gamma) \leq 2.7 \times 10 ^{-6}$ 
 \cite{Edwards:1996te},
imply that
\begin{equation}
R_\mu \leq 1.8 \times 10^{-6},\ R_e \leq 1.6 \times 10^{-5}
\end{equation}
and these in turn, can be used to place the constraints
\begin{eqnarray}
 \left|\sum_N \ln{\mu}\lns{\tau} F(\lambda_N)\right|^2 \leq 
1.2 \times 10^{-4}, &&
\left|\sum_N \ln{e} \lns{\tau} F(\lambda_N)\right|^2 \leq 
1.0 \times 10^{-3}.
\label{constraint}
\end{eqnarray}
If the mixing angles are such that only one heavy neutrino is 
important, we may use $F(x)\to -1/4$ as $x \to \infty$ to estimate that
\begin{eqnarray}
\left| \ln{\mu}\lns{\tau} \right|\leq 0.044, && \left| \ln{e} \lns{\tau}\right| \leq 0.13.
\end{eqnarray}
For comparison, the experimental bound $B(\mu \to e \gamma) < 1.2 \times 10^{-11}$  \cite{Eidelman:2004wy} leads to the constraint
\begin{eqnarray}
\left| \ln{e}\lns{\mu} \right|\leq 1.2\times 10^{-4}.
\label{muegb}
\end{eqnarray}
Similarly, for $W^{\prime}$ exchange one obtains
\begin{equation}
\Gamma(\ell^\prime \rightarrow \ell \gamma) = {G_F^2 \alpha \over
32 \pi^4} m_{\ell^{\prime}}^5 \left| \sum_N \rn{\ell} \rns{\ell^\prime}
\beta F(\beta \lambda_N)\right|^2.
\label{lptolgr}
\end{equation}
Considering once again the case where only one heavy neutrino comes into play, and with $\lambda_{N}\to \infty$ this leads to
\begin{eqnarray}
\left| \rn{\mu}\rns{\tau} \right|\leq \frac{0.044}{\beta}, && \left| \rn{e} \rns{\tau}\right| \leq \frac{0.13}{\beta}.
\end{eqnarray}
For a typical $\beta \sim 0.01$ these limits are about an order of magnitude worse than the unitarity constraints. The limits become weaker by a factor of four for $M_{N}\sim M_{R}$.

If the mixing parameter $\xi_{W}$ is not zero, the second operator in Eq.~\ref{lplgops} can dominate the rate because it is not proportional to the light lepton mass. In this case we obtain 
\begin{eqnarray}
\Gamma(\ell^\prime \rightarrow \ell \gamma)& =& {G_F^2 \alpha \over
32 \pi^4} m_{\ell^{\prime}}^3 M_W^2 \xi_g^2 \left(
\left|\sum_N \rn{\ell} \lns{\ell^\prime}
\sqrt{\lambda_N}\tilde F(\lambda_N)\right|^2 \right.\nonumber \\
&+&\left. \left|\sum_N \ln{\ell} \rns{\ell^\prime}
\sqrt{\lambda_N}\tilde F(\lambda_N)\right|^2 \right) .
\label{lptolgmix}
\end{eqnarray}
With only one heavy neutrino, and using $\tilde F(x) \to -1/4$ as 
$x\to \infty$, the constraints are
\begin{eqnarray}
\xi_g^2 \lambda_{N} \left(\rmu\lt+\lmu\rt \right)  
&\leq & 9.3 \times 10^{-7},
\nonumber \\
\xi_g^2 \lambda_{N} \left(\re\lt+\le\rt \right)  
&\leq & 8.0 \times 10^{-6}, \nonumber \\
\xi_g^2 \lambda_{N} \left(\re\lmu+\le\rmu \right)  
&\leq & 2.4 \times 10^{-14}.
\label{constraintmix}
\end{eqnarray}
The last result follows from the corresponding analysis for $\mu \to e \gamma$.  For LR models with $W_{L}-W_{R}$ mixing,  
LFV $B$ decay modes are proportional to 
$\xi_{g}^{4}$ making Eq.~\ref{constraintmix} the most stringent 
constraint in this case. 

\section{LFV $\tau$ decay into three charged leptons}\label{s:t3l}

The pure radiative decays discussed so far do not constrain 
LR models without  $W_{L}-W_{R}$ mixing  because  
there is no $W^\pm_L W^\mp_R \gamma$ vertex. Similarly, there is no
$W^\pm_L W^\mp_R Z$ vertex, and the modes 
$\tau^- \rightarrow \ell_1^- \ell_2^- \ell_3^+$ proceed through box diagrams at leading order. We now derive the constraints  that can be placed on the neutrino mixing matrix from these modes.

\begin{figure}[!htb]
\begin{center}
\includegraphics[width=6cm]{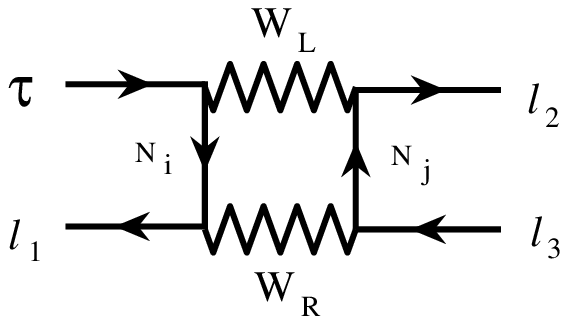}
\end{center}
\caption{Box diagram responsible for the decays
$\tau^- \rightarrow \ell_1^- \ell_2^- \ell_3^+$ in models with no
$W_L-W_R$ mixing.}
\label{f:boxl}
\end{figure}

The effective operator responsible for these decay modes can be calculated from the diagram in Figure~\ref{f:boxl} 
plus two other diagrams obtained by interchanging $W_L \leftrightarrow W_R$
and by interchanging $\ell_1 \leftrightarrow \ell_2$.
Using dimensional regularization we find
\begin{eqnarray}
{\cal L} &=& {G_F \over \sqrt{2}} {\alpha \over 2 \pi
\sin^2\theta_W} \beta_g \sum_{N_i,N_j} \left\{E^S_{LR}(\lambda_{N_i},\lambda_{N_j},\beta)  \right.\nonumber \\
&& \left[
\lnis{\tau}{i} \rni{\ell_1}{i} \lni{\ell_2}{j} \rnis{\ell_3}{j}
\bar\ell_2 P_R \ell_3 \bar \ell_1 P_L \tau +
\rnis{\tau}{i} \lni{\ell_1}{i}\rni{\ell_2}{j} \lnis{\ell_3}{j}
\bar\ell_2 P_L \ell_3 \bar \ell_1 P_R \tau \right], \nonumber \\
&+&
E^T_{LR}(\lambda_{N_i},\lambda_{N_j},\beta) \left[
\lnis{\tau}{i} \rni{\ell_1}{i} \lni{\ell_2}{j}\rnis{\ell_3}{j}
\bar\ell_2 \gamma_\nu \gamma_\mu P_R \ell_3 \bar \ell_1 \gamma_\mu \gamma_\nu
P_L \tau \right . \nonumber \\
& + & \left.\left . \rnis{\tau}{i} \lni{\ell_1}{i} \rni{\ell_2}{j}\lnis{\ell_3}{j}
\bar\ell_2 \gamma_\nu \gamma_\mu P_L \ell_3 \bar \ell_1 \gamma_\mu \gamma_\nu P_R \tau
\right] ~+~ (\ell_1 \leftrightarrow \ell_2) \right\},
\label{t3lres}
\end{eqnarray}
where the Inami-Lim functions $E^{S,T}_{LR}$ are given by
\begin{eqnarray}
 E^S_{LR}(\lambda_{N_i}, \lambda_{N_j}, \beta) &=& \sqrt{\lambda_{N_i} \lambda_{N_j}}\left[
\frac{1}{\hat\epsilon} + \log\left({\mu^2\over M_W^2}\right) + 1 +
\frac{ \log\beta}{(1-\beta)(1-\beta \lambda_{N_i})(1-\beta \lambda_{N_j})}
\right.\nonumber \\
& + & 
\frac{\beta \lambda_{N_i}^3
\log \lambda_{N_i}}{(\lambda_{N_j}-\lambda_{N_i})(1-\lambda_{N_i})(1-\beta \lambda_{N_i})}\nonumber \\
& + & \left.
\frac{\beta \lambda_{N_j}^3 \log \lambda_{N_j}}{(\lambda_{N_i}-\lambda_{N_j})
(1-\lambda_{N_j})(1-\beta \lambda_{N_j})} \right], \nonumber \\ 
E^T_{LR}(\lambda_{N_i}, \lambda_{N_j},\beta) &= & \sqrt{\lambda_{N_i} \lambda_{N_j}}\left[
\frac{\lambda_{N_i} \log \lambda_{N_i}}  
{(\lambda_{N_j}-\lambda_{N_i})
(1-\lambda_{N_i})(1-\beta \lambda_{N_i})}
\left(1-\frac{\lambda_{N_i}}{4}(1+\beta)\right) 
%-\frac{1+\beta}{4}\frac{ \lambda_{N_i}^2 \log \lambda_{N_i}}
%{(\lambda_{N_j}-\lambda_{N_i})(1-\lambda_{N_i})(1-\beta %\lambda_{N_i})}  
\right. \nonumber \\
&+&  \frac{\lambda_{N_j} \log \lambda_{N_j}}{(\lambda_{N_i}-\lambda_{N_j})(1-\lambda_{N_j})
(1-\beta \lambda_{N_j})}
\left(1-\frac{\lambda_{N_j}}{4}(1+\beta)\right) 
%-\frac{1+\beta}{4}\frac{ \lambda_{N_j}^2 \log \lambda_{N_j}}
%{(\lambda_{N_i}-\lambda_{N_j})(1-\lambda_{N_j})(1-\beta %\lambda_{N_j})}   
\nonumber \\
& +& \left. \frac{\left(3\beta -1\right)\log \beta}{4(1-\beta)(1-\beta \lambda_{N_i})(1-\beta \lambda_{N_j})}
%-\frac{1+\beta}{4}\frac{\log\beta}{(1-\beta)(1-\beta \lambda_{N_i})(1-%\beta \lambda_{N_j})}
\right].
\label{tensorfflr}
\end{eqnarray}
This result is divergent and we have regulated the 
divergence by defining
\begin{equation}
{1\over \hat\epsilon} = {2 \over 4-n} +\log 4\pi-\gamma.
\end{equation}
The divergence arises because in LR models these box 
diagrams are not the only ones that contribute to this process. 
In particular, these models require the existence 
of neutral scalars with tree-level flavor changing couplings \cite{probslr}. These scalars give rise to a tree-level amplitude 
for this process as well as to several additional one-loop diagrams. 
To keep our analysis as model independent and simple as possible, we will not specify the scalar sector of the left-right models. Instead we will use Eq.~\ref{tensorfflr}, drop the $1/\hat\epsilon$ pole, and take a 
scale $\mu \sim 1$~TeV. This approach can be considered as a limit in which the scalars that make the left-right model renormalizable are very heavy. To gain some insight into our prescription, we compare 
our result to the complete calculation of Ref.~\cite{pilaftsis} for 
$K_{L}\to \mu e$ in the appendix. 

Taking both the muon and electron to be massless, we can calculate the rates:
\begin{eqnarray}
\Gamma(\tau &\rightarrow &\ell_1^- \ell_2^- \ell_3^+) \ =\  {G_F^2 m_\tau^5 \over 192
\pi^3} \left({\alpha^2 \over 128 \pi^2 \sin^4\theta_W}\right)
\beta_g^2  \left. \{ \right . \nonumber \\
&& \left . \left|\sum_{N_i N_j} \lnis{\tau}{i} \rni{\ell_1}{i} \lni{\ell_2}{j}\rnis{\ell_3}{j} E^S_{LR}\right|^2 + 64 \left|\sum_{N_i N_j}  \lnis{\tau}{i} \rni{\ell_1}{i} \lni{\ell_2}{j}\rnis{\ell_3}{j} E^T_{LR}\right|^2 \right. \nonumber \\
& + &  \left. 8 {\it Re}\left[ \left(\sum_{N_i N_j} \lnis{\tau}{i} \rni{\ell_1}{i} \lni{\ell_2}{j}\rnis{\ell_3}{j} E^S_{LR}\right)\left(\sum_{N_i N_j}  \lni{\tau}{i} \rnis{\ell_1}{i} \lnis{\ell_2}{j}\rni{\ell_3}{j} E^T_{LR}\right)\right] \right. \nonumber \\
&+& \left .\left|\sum_{N_i N_j}\rnis{\tau}{i} \lni{\ell_1}{i} \rni{\ell_2}{j}\lnis{\ell_3}{j}E^S_{LR}\right|^2  + 64 \left|\sum_{N_i N_j}\rnis{\tau}{i} \lni{\ell_1}{i} \rni{\ell_2}{j}\lnis{\ell_3}{j}E^T_{LR}\right|^2  \right .
\nonumber \\
& + & \left. 8{\it Re}\left[\left(\sum_{N_i N_j}\rnis{\tau}{i} \lni{\ell_1}{i} \rni{\ell_2}{j}\lnis{\ell_3}{j}E^S_{LR}\right)\left( \sum_{N_i N_j}\rni{\tau}{i} \lnis{\ell_1}{i} \rnis{\ell_2}{j}\lni{\ell_3}{j}E^T_{LR}\right)\right] \right . \nonumber \\
&+& \left . (\ell_1 \leftrightarrow \ell_2) \right \}.
\end{eqnarray}

Before comparing with experiment it is convenient to define the ratio 
\begin{equation}
R_{123}(\tau \rightarrow \ell_1^- \ell_2^- \ell_3^+) \equiv
{\Gamma (\tau \rightarrow \ell_1^- \ell_2^- \ell_3^+) \over
\Gamma(\tau^- \rightarrow \mu^- \nu_\tau \bar\nu_\mu)},
\end{equation}
and normalize the rates this way. When 
only one heavy neutrino $N$ is important, this simplifies to
\begin{eqnarray}
R_{123}(\tau &\rightarrow& \ell_1^- \ell_2^- \ell_3^+) \ =\ 
\left(|\lns{\tau} \rn{\ell_1}|^2 | \ln{\ell_2} \rns{\ell_3}|^2 +
|\rns{\tau}\ln{\ell_1}|^2 |\rn{\ell_2}\lns{\ell_3} |^2  \right .
\nonumber \\
& + & \left . |\lns{\tau}\rn{\ell_2}|^2 |\ln{\ell_1}\rns{\ell_3}|^2 + |\rns{\tau}\ln{\ell_2}|^2 |\rn{\ell_1}\lns{\ell_3} |^2 \right ) F_{LR}(\lambda_N,\beta),
\end{eqnarray}
where we have defined the form factor 
\begin{eqnarray}
F_{LR}(\lambda_N,\beta) &\equiv &
{\alpha^2 \beta_g^2\over 128 \pi^2 \sin^4\theta_W}  
\left[ E^S_{LR}(\lambda_N,\lambda_N,\beta)^2 +
8 E^S_{LR}(\lambda_N,\lambda_N,\beta) E^T_{LR}(\lambda_N,\lambda_N,\beta)\right. \nonumber \\
&+&\left.
64 E^T_{LR}(\lambda_N,\lambda_N,\beta)^2 \right].
\end{eqnarray}

\begin{figure}[htbp]
\begin{center}
\includegraphics[width=8cm]{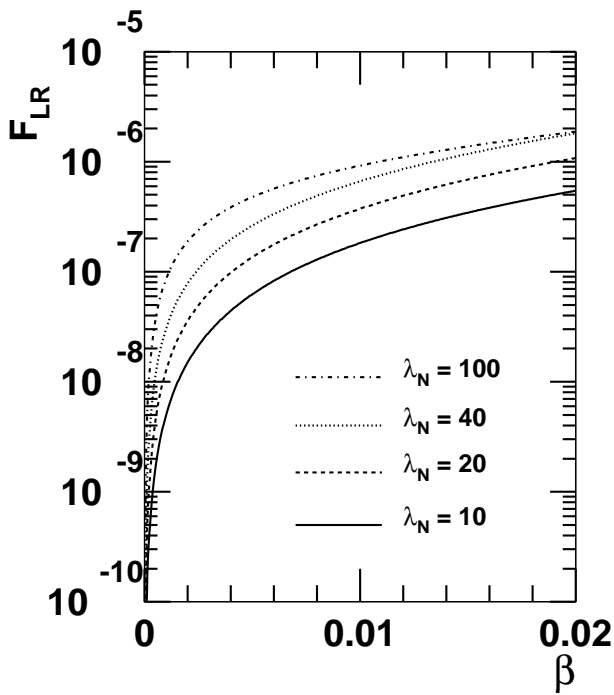}
\caption{$F_{LR}(\lambda_N,\beta)$ with $g_{R}=g_{L}$ as a function of $\beta$ and 
$\lambda_{N}$.}
\end{center}
\label{f:tauflr}
\end{figure}

With the experimental limits from Ref.~ \cite{Aubert:2003pc,Yusa:2004gm} we find
\begin{eqnarray}
R_{e\mu\mu} &=& F_{LR}(\lambda_N,\beta) (\lt\rmu +\rt\lmu)(\lmu\re+\le\rmu)   <  1.2 \times 10^{-6} \nonumber \\
R_{\mu ee} &=& F_{LR}(\lambda_N,\beta) (\lt\re +\rt\le)(\lmu\re+\le\rmu)   < 1.1 \times 10^{-6} \nonumber\\
R_{\mu\mu e} &=& 2 F_{LR}(\lambda_N,\beta) \lmu\rmu(\lt\re+\rt\le)   <  1.2 \times 10^{-6} \nonumber\\
R_{ee\mu} &=& 2F_{LR}(\lambda_N,\beta) \le\re(\lt\rmu+\rt\lmu)  <  1.2 \times 10^{-6}  \nonumber\\
R_{\mu\mu\mu} &=& F_{LR}(\lambda_N,\beta) \lmu\rmu(\lt\rmu+\rt\lmu)  <   1.2 \times 10^{-6} \nonumber\\
R_{eee} &=& F_{LR}(\lambda_N,\beta) \le\re(\lt\re+\rt\le)  < 2.0 \times 10^{-6}
\label{tto3lconstraints}
\end{eqnarray}
In Figure~\ref{f:tauflr} we show the value of $F_{LR}(\lambda_N,\beta)$ for $g_{R}=g_{L}$ 
as a function of $\beta$ for selected values of $\lambda_{N}$. 
This figure indicates that for a wide range of 
parameters, $F_{LR}(\lambda_N,\beta)$  is between $10^{-7}$ and 
$10^{-6}$. With this range in Eq.~\ref{tto3lconstraints} we can see that the present experimental limits on $R_{123}$ do not yield constraints on the mixing parameters that are significantly better than the unitarity bounds.

A similar exercise for muon decay yields  \cite{Eidelman:2004wy}
\begin{eqnarray}
\frac{\Gamma(\mu \rightarrow e e e)}{\Gamma(\mu \rightarrow e \nu_\mu \bar \nu_e)} = F_{LR}(\lambda_N,\beta) \le\re (\lmu\re+\le\rmu) 
\leq 1.0 \times 10^{{-12}}.
\end{eqnarray}
In this case, if we assume that all the angles are of the same order, we obtain the constraint $U_{ij}^{LR} \leq 0.2$. The $\tau$ decay modes are still far from achieving this level of sensitivity. If we write for example,
\begin{eqnarray}
R_{\mu ee} &=& \left(\left|\frac{\ln{\tau}}{\ln{e}}\right|^2 +\left|\frac{\rn{\tau}}{\rn{e}}\right|^2\right) \frac{\Gamma(\mu \rightarrow e e e)}{\Gamma(\mu \rightarrow e \nu_\mu \bar \nu_e)} \nonumber\\ 
&\leq & 1.0 \times 10^{{-12}}\left(\left|\frac{\ln{\tau}}{\ln{e}}\right|^2 +
\left|\frac{\rn{\tau}}{\rn{e}}\right|^2\right) ,
\end{eqnarray}
we see that the $\tau$ decay bounds need to improve by six orders of magnitude in order to be competitive with the already available muon decay limit. Of course, the muon decay is not sensitive to all the parameters needed to describe neutrino mixing in general and the $\tau$ decay data is complementary.

\section{LFV $B$ decay: operators}

\begin{figure}[!htb]
\begin{center}
\includegraphics[width=6cm]{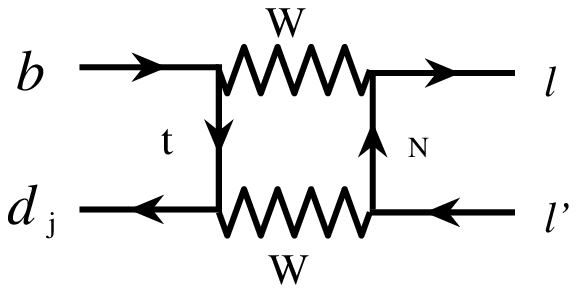}
\end{center}
\caption{Box diagram responsible for the process $b\to d_j \bar \ell \ell^\prime$. }
\label{f:box}
\end{figure}

We now turn our attention to $B$ decay modes and start by calculating the basic quark level LFV process $b\to d_j \bar \ell \ell^\prime$. In unitary gauge the process occurs through the 
box diagram of Figure~\ref{f:box} . 
We distinguish several cases as before: left-handed heavy neutrinos; 
LR models with $W_{L}-W_{R}$ mixing; right-handed heavy neutrinos; 
and LR  models without $W_{L}-W_{R}$ mixing.

\subsection{Left-handed heavy neutrinos}

A straightforward calculation of the diagram in Figure~\ref{f:box} produces the operator:
\begin{equation}
{\cal L}= {G_F \over \sqrt{2}} {\alpha \over 8 \pi
\sin^2\theta_W}
V^{L*}_{td_j}V^L_{tb} \ln{\ell} \lns{\ell^\prime}
E_L(\lambda_t,\lambda_N) \bar\ell \gamma_\mu
P_L \ell^\prime
\bar d_j \gamma^\mu P_L  b,
\label{bleftnu}
\end{equation}
where $d_j$ refers to a $d$ or an $s$ quark.
With the aid of the unitarity relations of Eq.~\ref{unitarity},
we find
\begin{eqnarray}
E_L(\lambda_t,\lambda_N) &=& \lambda_t \lambda_N
\left[ {3\over (1-\lambda_N)(1-\lambda_t)} \right. \nonumber \\
&+& \left.
{\left(4-8\lambda_t+\lambda_t^2\right)\log\lambda_t \over
(\lambda_N-\lambda_t)(1-\lambda_t)^2}+
{\left(4-8\lambda_N+\lambda_N^2\right)\log\lambda_N \over
(\lambda_t-\lambda_N)(1-\lambda_N)^2} \right].
\label{boxl}
\end{eqnarray}
This result is in agreement with the existing result for 
$K_{L}\rightarrow \mu^{\pm} e^{\mp}$  \cite{pilaftsis} when the 
$b$ quark is replaced by an $s$ quark.

\subsection{Left-right model with $W_L-W_R$ mixing}

In LR models with mixing there is a second operator that can be obtained from Figure~\ref{f:box},
\begin{eqnarray}
{\cal L}&=& {G_F \over \sqrt{2}} {\alpha \over 2 \pi
\sin^2\theta_W} \xi_g^2 \nonumber \\
&\times& \left[ \tilde{E}^S_L(\lambda_t,\lambda_N)
\bar\ell \left(\lns{\ell} \rn{\ell^\prime} P_R + \rns{\ell} \ln{\ell^\prime} P_L\right) \ell^\prime
\bar d_j \left(V^{L*}_{td_j}V^R_{tb}P_R + V^{R*}_{td_j}V^L_{tb}P_L \right) b \right.\nonumber \\
&+& \left.\tilde{E}^T_L(\lambda_t,\lambda_N)
\bar\ell \gamma_\nu \gamma_\mu \left(\lns{\ell} \rn{\ell^\prime} P_R + \rns{\ell} \ln{\ell^\prime} P_L\right)\ell^\prime
\bar d_j \gamma^\mu \gamma^\nu \left(V^{L*}_{td_j}V^R_{tb}P_R + V^{R*}_{td_j}V^L_{tb}P_L \right) b \right ]. \nonumber \\
\label{bopmix}
\end{eqnarray}
The Inami-Lim functions are the same as those in Eq.~\ref{tensorfflr} with $\beta=1$. As mentioned above, these operators produce observables proportional to $\xi_{g}^{4}$ and cannot place constraints that are competitive with LFV radiative $\tau$ decay.

\subsection{Right-handed heavy neutrino}

Models with a mostly right handed heavy neutrino would proceed 
through the diagram in Figure~\ref{f:box} with two $W^{\prime}$ bosons exchanged. Ignoring the $W_{L}-W_{R}$ mixing this 
results in an operator
\begin{equation}
{\cal L}=\beta {G_F \over \sqrt{2}} {\alpha \over 8 \pi
\sin^2\theta_W}
V^{R*}_{td_j}V^R_{tb} \rn{\ell} \rns{\ell^\prime}
E_L(\beta\lambda_t,\beta\lambda_N) \bar\ell \gamma_\mu
P_R \ell^\prime
\bar d_j \gamma^\mu P_R  b
\label{brightnu}
\end{equation}
A glance at Eq.~\ref{boxl} reveals that this operator will make contributions to LFV $B$ decay rates that are suppressed by at least a factor of $\beta^{4}$.

\subsection{Left-right models without $W_{L}-W_{R}$ mixing}

In this scenario the heavy neutrinos have left and right-handed
couplings and both the $W$ and $W^\prime$ appear. 
The LFV operator for $B$ decay 
arises from box diagrams like the one in Figure~\ref{f:box} with one $W$ and one $W^\prime$, we find
\begin{eqnarray}
{\cal L}&=& {G_F \over \sqrt{2}}  {\alpha \over 2 \pi
\sin^2\theta_W}\beta_g
\left[  E^S_{LR}(\lambda_t,\lambda_N, \beta)
\left( V^{L*}_{td_j} V^R_{tb}\rn{\ell} \lns{\ell^\prime}
\bar \ell P_L \ell^\prime \bar
d_j  P_R b \right . \right. \nonumber \\
&+& \left.
V^{R*}_{td_j} V^L_{tb} \ln{\ell} \rns{\ell^\prime} \bar \ell P_R
\ell^\prime \bar
d_j  P_L b \right)
 +  E^T_{LR}(\lambda_t,\lambda_N, \beta)
\left( V^{L*}_{td_j} V^R_{tb} \rn{\ell} \lns{\ell^\prime}  \bar
\ell \gamma_\nu\gamma_\mu P_L \ell^\prime
\bar d_j \gamma^\mu \gamma^\nu P_R b \right. \nonumber \\
&+& \left. \left. V^{R*}_{td_j} V^L_{tb} \ln{\ell} \rns{\ell^\prime} \bar \ell \gamma_\nu\gamma_\mu P_R \ell^\prime
\bar d_j  \gamma_\mu\gamma_\nu P_L b \right) \right],
\label{rl}
\end{eqnarray}
where the Inami-Lim functions were given in Eq.~\ref{tensorfflr}. 

For the process $B \rightarrow \bar{\ell} \ell^\prime$ the two operators in Eq.~\ref{rl} can be reduced to one by using the relation
\begin{equation}
\gamma_\nu \gamma_\mu \otimes \gamma_\mu \gamma_\nu \ \rightarrow
\ 4( 1 \otimes 1) + \sigma_{\mu\nu} \otimes  \sigma_{\mu\nu},
\label{diracsimp}
\end{equation}
and dropping the last term in anticipation of the vanishing of the matrix element
\begin{equation}
<0|\bar{d}_j\sigma_{\mu\nu}(1\pm\gamma_5)b|\bar B^0_j> = 0.
\end{equation}
We obtain,
\begin{eqnarray}
{\cal L} &=& {G_F \over \sqrt{2}} {\alpha \over 2 \pi
\sin^2\theta_W} \beta_g 
\left(E^S_{LR}(\lambda_t,\lambda_N, \beta) + 4 
E^T_{LR}(\lambda_t,\lambda_N, \beta) \right)
\nonumber \\ &\times &
\left( V^{L*}_{td_j} V^R_{tb} \rn{\ell} \lns{\ell^\prime}  \bar \ell P_L \ell^\prime \bar
d_j  P_R b + V^{R*}_{td_j} V^L_{tb} \ln{\ell} \rns{\ell^\prime} \bar \ell P_R
\ell^\prime \bar d_j  P_L b \right).
\label{rl-1}
\end{eqnarray}

\section{LFV $B$ decay phenomenology}

We now use the operators obtained in the previous section to 
compute their contribution to selected LFV $B$ decay modes. 

\subsection{$B_{d_j} \rightarrow \tau^\pm \ell^\mp$}

We first consider the mode $B \rightarrow \tau^\pm \ell^\mp$, 
the analogue of the $K_{L}\to \mu^{\pm} e^{\mp}$ mode 
which has been discussed extensively in the literature  \cite{kmue,pilaftsis}. 

For a heavy left-handed neutrino we find after summing the two modes 
and neglecting the mass of $\ell =\mu,e$,
\begin{eqnarray}
\Gamma(B_j \rightarrow \tau^\pm \ell^\mp) &=& {1 \over 256}
{G_F^2 \over \pi} \left({\alpha  \over 4 \pi \sin^2\theta_W}\right)^2
F_B^2 m_\tau^2 M_B \left(1-{m_\tau^2 \over M_B^2}\right)^2 \nonumber \\
&& |V^{L\star}_{tb}V^L_{tj}|^2 \left|\sum_N \ln{\ell}\lns{\tau}
E_L(\lambda_t,\lambda_N)\right|^2.
\label{btotm}
\end{eqnarray}

When the heavy neutrino has both left and right handed couplings (with an explicit $W^{\prime}$ and no mixing) we find 
\begin{eqnarray}
\Gamma(B_j & \rightarrow &\tau^\pm \ell^\mp) \ =\ {1\over 32}
{G_F^2 \over \pi} \left({\alpha  \over 4 \pi \sin^2\theta_W}\right)^2
F_B^2  {M_B^5 \over m_b^2} \beta_g^2
 \left(1-{m_\tau^2 \over M_B^2}\right)^2 \nonumber \\
&&\left[ \left|\sum_N
V^{L\star}_{tb}V^R_{tj} \rn{\ell}\lns{\tau}\left[
E^{S}_{LR}(\lambda_t,\lambda_N, \beta)+ 4
E^{T}_{LR}(\lambda_t,\lambda_N, \beta)
\right]\right|^2  \right. \nonumber \\
&+& \left .\left|\sum_N V^{R\star}_{tb}V^L_{tj}\ln{\ell}\rns{\tau} \left[
E^{S}_{LR}(\lambda_t,\lambda_N, \beta)+ 4
E^{T}_{LR}(\lambda_t,\lambda_N, \beta)
\right]\right|^2\right].
\label{fullff}
\end{eqnarray}

For numerical purposes it is natural to compare these rates to the standard model rate for $B_{d_j}\rightarrow \tau^+\tau^-$,
\begin{equation}
\Gamma(B_j \rightarrow \tau^+ \tau^-) =
{G_F^2 \over \pi} \left({\alpha  \over 4 \pi \sin^2\theta_W}\right)^2
F_B^2 m_\tau^2 M_B \sqrt{1-{4m_\tau^2 \over M_B^2}}
|V^\star_{tb}V_{tj}|^2 Y^2(\lambda_t)
\label{bttsm}
\end{equation}
where the Inami-Lim function $Y(\lambda_t)\sim 1.06$  \cite{Buchalla:1995vs}.
We define
\begin{equation}
R_{B\tau \ell} \equiv {\Gamma(B_j \rightarrow \tau^\pm \ell^\mp)
\over \Gamma(B_j \rightarrow \tau^+ \tau^-)},
\end{equation}
and in terms of this definition we find for left handed heavy neutrinos
\begin{eqnarray}
R_{B\tau \ell} &=& 3.7 \times 10^{-3} \left| \sum_N \ln{\ell}\lns{\tau}E_L(\lambda_t,\lambda_N)\right|^2.
\label{auxl}
\end{eqnarray}

\begin{figure}[!htb]
\begin{center}
\includegraphics[width=7cm]{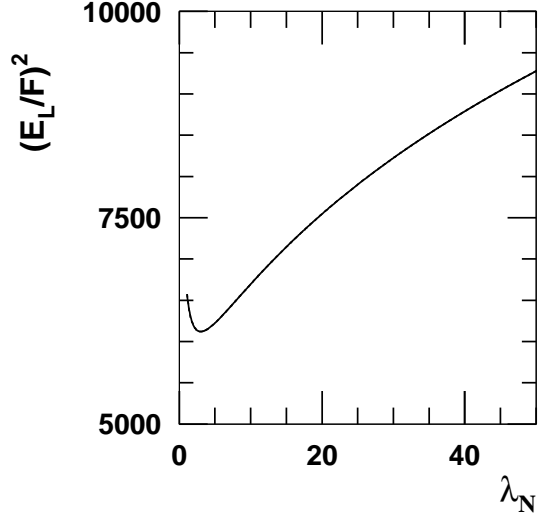}
\end{center}
\caption{Ratio of form factors 
$(E_L/F)^2$ as a function of $\lambda_N$.}
\label{f:left}
\end{figure}

To compare the sensitivity of the modes $B\to \tau^{\pm}\ell^{\mp}$ and $\tau \to \ell \gamma$ to the neutrino mixing parameters 
we plot in Figure~\ref{f:left}
the ratio $(E_L/F)^2$
as a function of $\lambda_N$.

Assuming there is only one heavy neutrino and using Eq.~\ref{constraint} as well as $(E_L/F)^2\sim 7500$ from Figure~\ref{f:left} we find
\begin{eqnarray}
R_{B\tau \mu} \leq 3.3 \times 10^{-3}, &&
R_{B\tau e} \leq  2.8 \times 10^{-2}.
\label{lbcomp}
\end{eqnarray}
The standard model expectations for $B_{d_j}\rightarrow \tau^+\tau^-$  \cite{Buchalla:1995vs} with 
the central values for the CKM angles found in  \cite{Battaglia:2003in} are 
\begin{eqnarray}
B(B_s \rightarrow \tau^+ \tau^-) = 1.1 \times 10^{-6}, &&
B(B_d \rightarrow \tau^+ \tau^-) = 3.3 \times 10^{-8}.
\label{bllsm}
\end{eqnarray}
Consequently, one would need a single event sensitivity of at least $10^{-8}$ for 
$B(B_{s}\to \tau^{\pm}e^{\mp})$  ($10^{-9}$ for $B(B_{s}\to \tau^{\pm}\mu^{\mp})$) to improve on the existing constraints from radiative $\tau$ decay.  
\begin{table}[htdp]
\caption{Summary of current experimental bounds for $B\to \ell^{+}\ell^{\prime -}$.} 
\begin{center}
\begin{tabular}{|c|c|}\hline
& Branching Ratio   \\ 
\hline 
$B\rightarrow e^\pm\mu^\mp$& $ <1.8\times 10^{-7}$ Babar~ \cite{Aubert:2004gm}\\  
&$ < 1.7\times 10^{-7}$ Belle~ \cite{Chang:2003yy} \\
$B\rightarrow e^\pm\tau^\mp$&  $ < 1.1\times 10^{-4}$ Cleo~ \cite{Bornheim:2004rx} \\
$B\rightarrow \mu^\pm\tau^\mp$&  $ < 3.8\times 10^{-5}$ Cleo~ \cite{Bornheim:2004rx} \\
\hline
\end{tabular}
\end{center}
\label{t:bres2m}
\end{table}
A glance at Table~\ref{t:bres2m} indicates that one would need an  order of magnitude improvement over the current best limit from Belle 
for $B\rightarrow e^\pm\mu^\mp$. 
There is some hope that this sensitivity may be attainable in the future. For example, the estimated single event sensitivity for the $B \to \mu \mu$ modes at CDF with $15~fb^{-1}$ is ~ \cite{Anikeev:2001rk}
\begin{eqnarray}
1.3 \times 10^{-9} & {\rm ~for~}&
B_s \rightarrow \mu^\pm \mu^\mp  
\nonumber \\
4.7 \times 10^{-10} & {\rm ~for~}&
B_d \rightarrow \mu^\pm \mu^\mp.
\end{eqnarray}
Of course, there are additional experimental difficulties for modes involving a $\tau$ lepton, but we may regard Eq.~\ref{lbcomp} as the benchmark needed to improve upon limits from radiative $\tau$ decay.
For comparison, the same analysis applied to the limit $B(K_{L}\to \mu^{\pm}e^{\mp}) < 4.7 \times 10 ^{-12}$  \cite{Eidelman:2004wy}, yields the bound 
\begin{eqnarray}
\left| \ln{e}\lns{\mu} \right|\leq 0.07,
\end{eqnarray}
which is also weaker than the corresponding bound from $\mu\to e \gamma$, Eq.~\ref{muegb}.

For right handed heavy neutrinos, Eq.~\ref{auxl} becomes
\begin{eqnarray}
R_{B\tau \ell} &=& 3.7 \times 10^{-3} \left| \sum_N \rn{\ell}\rns{\tau}\beta E_L(\beta\lambda_t,\beta\lambda_N)\right|^2.
\end{eqnarray}
The bounds obtained are worse than those for left-handed heavy 
neutrinos.

Finally for the case where the heavy neutrino has both left and right handed couplings and the $W_{L}-W_{R}$ mixing can be ignored we obtain 
\begin{eqnarray}
R_{B\tau \ell} &\sim &
0.36  \beta_g^2 {1\over |V^\star_{tb}V_{tj}|^2}
\nonumber \\
&& \left[\left|\sum_N V^{L\star}_{tb}V^R_{tj} \rn{\ell}\lns{\tau}
\left[E^S_{LR}(\lambda_t,\lambda_N, \beta) + 4 
E^T_{LR}(\lambda_t,\lambda_N, \beta) \right]
\right|^2 \right.\nonumber \\
&+&\left. \left|\sum_N V^{R\star}_{tb}V^L_{tj}\ln{\ell}\rns{\tau} 
\left[E^S_{LR}(\lambda_t,\lambda_N, \beta) + 4 
E^T_{LR}(\lambda_t,\lambda_N, \beta) \right]
\right|^2\right].
\label{rbtlte}
\end{eqnarray} 
\begin{figure}[htbp]
\begin{center}
\includegraphics[width=8cm]{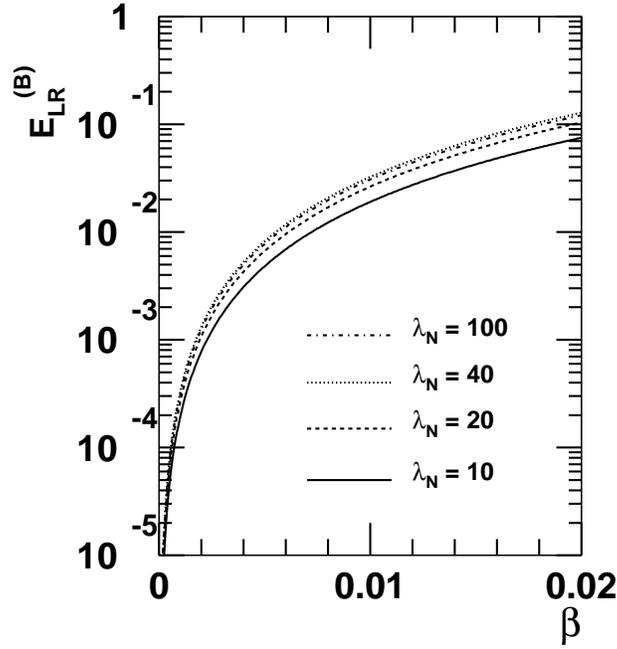}
\caption{$E^{(B)}_{LR}$ as a function of $\beta$ for 
selected values of $\lambda_{N}$.}
\end{center}
\label{fbtl}
\end{figure}
It is harder to interpret this result because there 
are several unknown parameters. To gain some insight into this 
result we consider the simplified case with  
only one heavy neutrino discussed in the introduction. We further assume that $V^R_{tj}\sim V^L_{tj}$, 
and  that $g_{L}=g_{R}$. We define the form factor
\begin{equation}
E^{(B)}_{LR} \equiv 0.36 \beta_{g}^{2}
\left[E^S_{LR}(\lambda_t,\lambda_N, \beta) + 4 
E^T_{LR}(\lambda_t,\lambda_N, \beta) \right]^{2},
\end{equation}
and plot it in Figure~6 as a function of $\beta$ for selected values of the heavy neutrino mass. We see from the 
figure that $E^{(B)}_{LR}$ can be between $0.01$ and $0.1$ for a wide range of parameters. Combining this result with Eq.~\ref{rbtlte} we find, 
\begin{equation}
R_{B\tau \ell} \sim (0.01-0.1)
\left[ |\rn{\ell}\lns{\tau}|^2 + |\ln{\ell}\rns{\tau}|^2\right].
\end{equation}
This in turn implies that $R_{B\tau \ell}$ has to be probed at the 
$10^{-3}$ level to constrain the neutrino mixing parameters beyond what is known from unitarity. Considering 
the SM expectation Eq.~\ref{bllsm},  
this implies a benchmark number for $B(B \rightarrow \tau^\pm \ell^\mp)$ at the $10^{-9}$ level, probably beyond reach for the foreseeable future. The corresponding bound 
$B(K_{L}\to \mu^{\pm}e^{\mp}) < 4.7 \times 10 ^{-12}$  \cite{Eidelman:2004wy} leads in turn to,
 \begin{equation}
\left[ |\rn{e}\lns{\mu}|^2 + |\ln{e}\rns{\mu}|^2\right] < 0.03.
\end{equation}

\begin{figure}[!htb]
\begin{center}
\includegraphics[width=10cm]{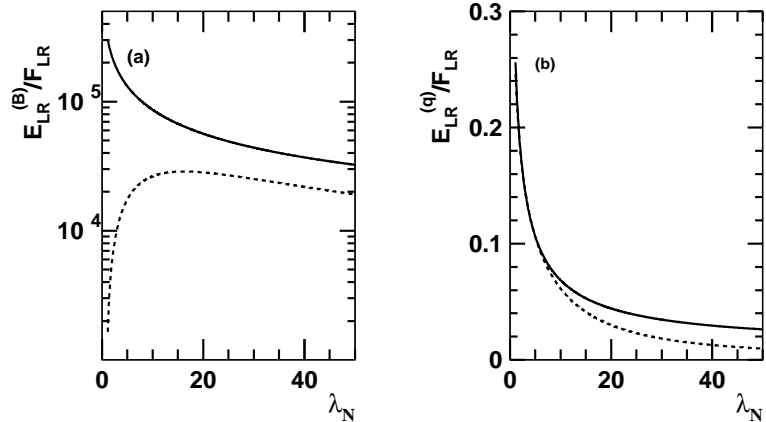}
\end{center}
\caption{Squared ratio of form factors for 
a) $E_{LR}^{(B)}/F_{LR}$ and b) $E_{LR}^{(q)}/F_{LR}$
as a function of $\lambda_N$ for $\beta= 0.0065$.}
\label{f:lrt3l}
\end{figure}

It is instructive to compare these modes to the 
$\tau^- \rightarrow \ell_1^- \ell_2^- \ell_3^+$ modes.  To this effect we 
plot in Figure~\ref{f:lrt3l}(a) the ratio of form factors $E_{LR}^{(B)}/F_{LR}$ as a 
function of $\lambda_N$ for $\beta= 0.0065$ (which corresponds to 
$M_{W^{\prime}}\sim 1$~TeV).
Using $E^{(B)}_{LR} \sim 10^{4} F_{LR}$  from Figure~\ref{f:lrt3l}~(a) we write
\begin{eqnarray}
B(B_{s}\to\tau^{\pm}\ell^{\mp}) \sim 0.01 F_{LR}
\left[ |\rn{\ell}\lns{\tau}|^2 + |\ln{\ell}\rns{\tau}|^2\right] 
\end{eqnarray}
which can be compared directly to Eq.~\ref{tto3lconstraints}. 

\subsection{Inclusive modes $b \rightarrow s \tau^\pm \ell^\mp$}

We turn our attention to the inclusive process $b \rightarrow s \tau^\pm \ell^\mp$. This choice is motivated by several factors. A 
semileptonic mode removes the helicity suppression present in the 
$B\to \tau^{\pm} \ell^{\mp}$ modes for the case of left-handed heavy 
neutrinos. \footnote{In the kaon sector the helicity suppression may be removed by considering instead $K\to \pi \mu e$ modes. The best bound in that case,  $\left| \ln{e}\lns{\mu} \right|\leq 2 $, arises from $B(K^{+}\to\pi^{+}\mu^{+}e^{-})<2.8 \times 10^{-11}$ \cite{Eidelman:2004wy} but is worse than the unitarity constraint.}
Requiring strangeness in the final state 
removes the CKM suppression factor $V_{td}/ V_{cb}$. Finally, 
considering an inclusive process removes the 
QCD suppression factor $F_B/M_B$. 

For left handed heavy neutrinos we find 
\begin{equation}
\Gamma(b \rightarrow d_j \tau^\pm \ell^\mp) =  {G_F^2 m_b^5 \over 192
\pi^3} \left({\alpha^2 \over 512 \pi^2 \sin^4\theta_W}\right)
I\left({m_\tau \over m_b}\right)
|V^{L\star}_{tb}V^L_{tj}|^2\left| \sum_N \ln{\ell}\lns{\tau} E_L(\lambda_t,\lambda_N)\right|^2,
\label{btostm}
\end{equation}
where
\begin{equation}
I(x)=1-8x^2+8x^6-x^8-24x^4\log(x)
\end{equation}
is the usual kinematic factor
for a non-zero $\tau$ mass and $I(m_\tau/m_b)\sim 0.33$ with
$m_b \sim 4.5$~GeV.
Using a $b$ lifetime given by 
\begin{equation}
\Gamma(b) = 5.8 {G_F^2m_b^5\over 192 \pi^3}
I\left({m_c \over m_b}\right)
 |V_{cb}|^2,
\label{gambq}
\end{equation}
with $I(m_c/m_b)\sim 0.5$ (for $m_c \sim 1.4$~GeV), this gives
\begin{eqnarray}
B(b \rightarrow d_j \tau^\pm \ell^\mp) & \approx &
2.6 \times 10^{-8} \left|{V_{tj}\over V_{cb}}\right|^2
\left|\sum_N \ln{\ell}\lns{\tau} E_L(\lambda_t,\lambda_N)\right|^2.
\end{eqnarray}
To compare with the radiative $\tau$ decay modes we assume that there is only one heavy neutrino, use
Eq.~\ref{constraint} as well as $|E_L/F|^2\sim 7500$ from Figure~\ref{f:left}~(a) to obtain,
\begin{eqnarray}
B(b \rightarrow s \tau^\pm \mu^\mp) &\leq & 2.3 \times 10^{-8} \nonumber \\
B(b \rightarrow s \tau^\pm e^\mp) &\leq & 2.0 \times 10^{-7} \nonumber \\
B(b \rightarrow d \tau^\pm \mu^\mp) &\leq & 9.1 \times 10^{-10} \nonumber \\
B(b \rightarrow d \tau^\pm e^\mp) &\leq & 7.7 \times 10^{-9}
\label{binccon}
\end{eqnarray}
These results imply that one needs at least a $10^{-7}$ sensitivity 
in $B(b \rightarrow s \tau^\pm e^\mp)$ to obtain constraints 
competitive with the radiative $\tau$ decay for left-handed heavy 
neutrinos.  Table~\ref{t:bres2} indicates that current B-factory 
results are close to this benchmark, although there are significant experimental hurdles for detection of $\tau$ leptons. 
\begin{table}[htdp]
\caption{Summary of current experimental bounds for $B \to X \ell^{+}\ell^{\prime -}$ modes.}
\begin{center}
\begin{tabular}{|c|c|}\hline
& Branching Ratio   \\ 
\hline 
$b \rightarrow s e^\pm \mu^\mp$ &$ < 2.2 \times 10^{-5}$ CLEO~ \cite{Glenn:1997gh} \\\hline
%$B \rightarrow X_s e^+e^- $ & $5.0 \times 10^{-6}$ %Belle~ \cite{Kaneko:2002mr}\\
%$B \rightarrow X_s \mu^+ \mu^-$ & $7.9 \times 10^{-6}$ %Belle~ \cite{Kaneko:2002mr}\\
%$B \rightarrow X_s \ell \ell$ & $5.6 \times 10^{-6}$ %BaBar~ \cite{Aubert:2004it} \\ 
%\hline
$B \rightarrow \pi e^\pm\mu^\mp$ & $ < 1.6 \times 10^{-6}$ CLEO~ \cite{Edwards:2002kq}  \\
$B \rightarrow K^0 e^\pm\mu^\mp$ & $ < 4.0 \times 10^{-6}$ BaBar~ \cite{Aubert:2002aw}  \\
$B \rightarrow K e^\pm\mu^\mp$ & $ < 1.6 \times 10^{-6}$ CLEO~ \cite{Edwards:2002kq}  \\
$B \rightarrow \rho e^\pm\mu^\mp$ & $ < 3.2 \times 10^{-6}$ CLEO~ \cite{Edwards:2002kq}  \\
$B \rightarrow K^*e^\pm \mu^\mp$ & $ < 3.4 \times 10^{-6}$  BaBar~ \cite{Aubert:2002aw}  \\ 
$B \rightarrow K^*e^\pm \mu^\mp$ & $ < 6.2 \times 10^{-6}$  CLEO~ \cite{Edwards:2002kq}  \\ 
\hline
%$B \rightarrow K e^\pm e^\mp$ & $ 4.8 \times 10^{-7}$ %Belle~ \cite{Ishikawa:2003cp}  \\
%$B \rightarrow K \mu^\pm \mu^\mp$ & $ 4.5 \times 10^{-7}$  %BaBar~ \cite{Aubert:2003cm}  \\
%$B \rightarrow K^* e^\pm e^\mp$ & $ 9.8 \times 10^{-7}$  %BaBar~ \cite{Aubert:2003cm}  \\
%$B \rightarrow K^* \mu^\pm \mu^\mp$ & $ 11.7 \times 10^{-7}$  %Belle~ \cite{Ishikawa:2003cp} \\
%\hline
\end{tabular}
\end{center}
\label{t:bres2}
\end{table}

As an estimate for the reach of future experiments we start from 
the Tevatron studies indicating that with $2~fb^{-1}$, CDF could detect 
61 $B_{d}\to K^{\star 0}\mu \mu$ events assuming a branching ratio 
of $1.5 \times 10^{-6}$  \cite{Anikeev:2001rk}. We can turn this number into an approximate single event sensitivity with $15 ~fb^{-1}$ of 
$3.2 \times 10^{-9}$ for this mode. This in turn indicates that improved constraints are possible, at least from the $b \to s \tau e$ mode.

\begin{figure}[!htb]
\begin{center}
\includegraphics[width=8cm]{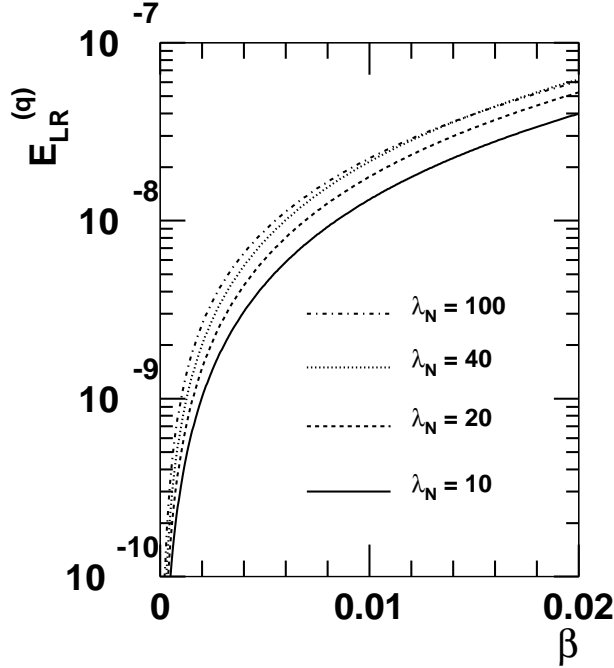}
\end{center}
\caption{$E^{(q)}_{LR}$ as a function of $\beta$ for selected values of $\lambda_{N}$ with $g_{R}= g_{L}$.}
\label{f:lrns}
\end{figure}

For heavy neutrinos with left and right handed couplings but vanishing $W_{L}-W_{R}$ mixing we find
\begin{eqnarray}
\Gamma(b \rightarrow d_j \tau^\pm \ell^\mp) &= &  {G_F^2 m_b^5 \over 192
\pi^3} \left({\alpha^2 \over 128 \pi^2 \sin^4\theta_W}\right)
\beta_g^2 I\left({m_\tau \over m_b}\right) \nonumber \\
&\times &\left(\left|\sum_N V^{L\star}_{tb}V^R_{tj} \rn{\ell}\lns{\tau}E^S_{LR}\right|^2 + 64\left|\sum_N V^{L\star}_{tb}V^R_{tj} \rn{\ell}\lns{\tau}E^T_{LR}\right|^2  \right. \nonumber \\
&+& \left. 8{\it Re}\left[\left(\sum_N V^{R\star}_{tb}V^L_{tj} \ln{\ell}\rns{\tau}
E^S_{LR}\right)\left( \sum_N V^{R}_{tb}V^{L\star}_{tj} \lns{\ell}\rn{\tau}
E^T_{LR}\right)\right]\right)
\end{eqnarray}
If we assume that only one heavy neutrino is important, 
that $g_{R}\sim g_{L}$, and  that $V^R_{tj}\sim V^L_{tj}$, we can write
\begin{eqnarray}
B(b \rightarrow d_j \tau^\pm \ell^\mp) =  
 \left|{V_{tj}\over V_{cb}}\right|^2
\left(\left| \rn{\ell}\lns{\tau}\right|^2
+ \left| \ln{\ell}\rns{\tau}\right|^2\right)
E^{(q)}_{LR}(\lambda_t,\lambda_N,\beta)
\end{eqnarray}
where we have defined
\begin{equation}
E^{(q)}_{LR}(\lambda_t,\lambda_N,\beta) \equiv
1.04 \times 10^{-7}\beta_{g}^{2} 
\left[(E^S_{LR})^2 +
8 E^S_{LR} E^T_{LR} +
64 ( E^T_{LR})^2\right].
\label{fflrq}
\end{equation}
From Figure~\ref{f:lrns} we see that $E^{(q)}_{LR}\sim 10^{-8}$ 
for a wide range of parameters. 
This allows us to write 
\begin{eqnarray}
B(b \rightarrow s \tau^\pm \ell^\mp) \sim 1 \times 10^{-8}
\left(\left| \rn{\ell}\lns{\tau}\right|^2
+ \left| \ln{\ell}\rns{\tau}\right|^2\right).
\end{eqnarray}
A sensitivity of at least $10^{-8}$ is thus required for the mode 
$b \rightarrow s \tau^\pm \ell^\mp$ to place significant constraints on the neutrino mixing parameters.

To compare these modes to the 
$\tau^- \rightarrow \ell_1^- \ell_2^- \ell_3^+$ modes, we 
plot in Figure~\ref{f:lrt3l}(b) the ratio of form factors $E_{LR}^{(q)}/F_{LR}$ as a 
function of $\lambda_N$ for $\beta= 0.0065$. 
Using $E^{(q)}_{LR} \sim 0.02 F_{LR}$ we write
\begin{eqnarray}
B(b \rightarrow s \tau^\pm \ell^\mp) \sim 0.02 F_{LR}
\left(\left| \rn{\ell}\lns{\tau}\right|^2
+ \left| \ln{\ell}\rns{\tau}\right|^2\right)
\end{eqnarray}
which can be compared directly to Eq.~\ref{tto3lconstraints}.  

\section{Summary and Conclusions}

We have studied LFV $\tau$ and $B$ decay modes within the 
context of neutrino mixing with additional heavy neutrinos. We 
have considered generic left-right models and distinguished three scenarios. In the first scenario we consider, the $W^{\prime}$ has a negligible effect and the lepton interactions are purely 
left-handed. In this case the best constraints on the parameters describing the neutrino mixing arise from radiative $\tau$ decay. 
Assuming one very heavy neutrino dominates, the current best bound 
is placed by $\tau \to \mu \gamma$,
\begin{equation}
|\ln{\mu}\lns{\tau}| \leq 0.044.
\end{equation}
To obtain a competitive constraint from $B_{s}\to \tau^{\pm}\mu^{\mp}$, this mode would have to be probed at the $10^{-9}$ level as 
indicated by Eqs.~\ref{lbcomp}~and~\ref{bllsm}. Similarly, to obtain 
a comparable constraint in any of the inclusive modes, 
$b \to s \tau^{\pm}e^{\mp}$ needs to be probed at the  $10^{-7}$ 
level as indicated in Eq.~\ref{binccon}.

The second scenario we considered involved a very heavy 
$W^{\prime}$. In this case the effect of the right-handed interaction can only be felt at low energies through $W_{L}-W_{R}$ mixing. 
In this scenario the best constraints arise from radiative $\tau$ decay 
and lead to unobservably small rates in LFV $B$ decay modes.

The third and final scenario we considered is one in which both the 
$W$ and the $W^{\prime}$ play a role in the lepton charged currents, but there is no $W_{L}-W_{R}$ mixing. In this more general case, the 
$(6\times 6)$ mixing matrix in the neutrino sector has many unknown parameters and as a practical matter different decay modes will in general probe different combinations of these parameters. As a benchmark for the sensitivity needed to probe this scenario we have considered three simplified cases with results summarized in 
Table~\ref{lrsumm}. The salient features are
\begin{table}[htdp]
\caption{Summary of results for a heavy neutrino with left and right handed couplings. We consider three cases with only one non-zero mixing with a heavy right-handed neutrino: a) $U^{R}_{eN}\neq 0$, 
b) $U^{R}_{\mu N}\neq 0$, and c) $U^{R}_{\tau N}\neq 0$.}
\begin{center}
\begin{tabular}{|c|c|c|c|}\hline
& $U^{R}_{eN}\neq 0$ &  $U^{R}_{\mu N}\neq 0$ & 
$U^{R}_{\tau N}\neq 0$ \\ [0.1ex]
\hline
$R_{e\mu\mu}/F_{LR}$ & 0 &  $\lt \le \rmus$  & 0 \\ [1ex]\hline
$R_{\mu ee}/F_{LR}$ &  $\lt\lmu\res$ & 0 & 0 \\[1ex]\hline
$R_{\mu\mu\mu}/F_{LR}$ &0& $\lmu \lt\rmus $ & 0 \\ [1ex]\hline
$R_{eee}/F_{LR}$&  $\le\lt\res$ &0  & 0 \\[1ex]
\hline
$R_{B\tau \mu}/E_{LR}^{(B)}$ & 0 & $\left|\rn{\mu}\right|^2\left|\ln{\tau}\right|^2$ &
2$\left|\rn{\tau}\right|^2\left|\ln{\tau}\right|^2$
\\[1.ex]\hline 
$R_{B\tau e}/E_{LR}^{(B)}$ & $\left|\rn{e}\right|^2 \left|\ln{\tau}\right|^2$ & 0 &
2$\left|\rn{\tau}\right|^2\left|\ln{\tau}\right|^2$
\\[1.ex]\hline 
$B(b \rightarrow s \tau_\pm \mu_\mp)/E_{LR}^{(q)}$ &  0& $\left|\rn{\mu}\right|^2\left|\ln{\tau}\right|^2$ &
$2\left|\rn{\tau}\right|^2\left|\ln{\tau}\right|^2$
\\[1.ex] \hline 
$B(b \rightarrow s \tau_\pm e_\mp)/E_{LR}^{(q)}$ &  $\left|\rn{\mu}\right|^2\left|\ln{\tau}\right|^2$& 0 &
$2\left|\rn{\tau}\right|^2\left|\ln{\tau}\right|^2$
\\[1.ex] \hline 
\end{tabular}
\end{center}
\label{lrsumm}
\end{table}%

\begin{itemize}

\item Radiative $\tau$ decay modes do not probe this scenario in the limit of no $W_{L}-W_{R}$ mixing. 

\item Unitarity implies that the matrix elements of the neutrino mixing matrix satisfy $|U^{L,R}_{\ell,N}|\leq 1$. To place significant constraints (better than the unitarity limit), $R_{123}$ has to be measured with a sensitivity of at least $10^{-8}$, two orders of magnitude better than current limits. 

\item The benchmark for significant constraints from $B$ decay  is a sensitivity of $10^{-9}$ for $B(B_{s} \to \tau^{\pm}\ell^{\mp})$ and 
of $10^{-8}$ for $B(b\to s \tau^{\pm}\ell^{\mp})$.

\end{itemize}
In all cases the sensitivity to the neutrino mixing parameters is much smaller than what already exists from the study of $\mu\to e \gamma$ and $K_{L}\to \mu e$ modes. However, LFV $\tau$ and $B$ decay modes offer an opportunity to complement those results by providing constraints on the $U_{\tau N}$ mixing angles.

\appendix

\section{Comparison with the $M_{H}\to \infty$ limit of Ref.~\cite{pilaftsis}.}

In this appendix we compare the form factor in $M \to \ell_{1}^{+}
\ell_{2}^{-}$ (for a spinless meson $M$) as obtained from Eq.~\ref{tensorfflr} and the complete one-loop calculation of 
Ref.~\cite{pilaftsis}. In \cite{pilaftsis} a LR model with a simple 
scalar sector is considered and a complete (gauge independent and finite) result for the form factor is obtained. In addition to the 
gauge boson contribution that we consider in this paper, the complete 
result depends on the parameters of the scalar sector, namely at least 
three scalar masses and scalar sector couplings. To compare this to 
our result we take the Higgs masses to infinity in the result of 
Ref.~\cite{pilaftsis}.
 
In what follows we use the notation in Appendix-B of \cite{pilaftsis}. 
We drop a common factor containing the mixing angles, the coupling constants and the factor $\sqrt{\lambda_{i}\lambda_N}$. 
The diagrams that contribute to the form factor in the limit when 
the Higgs masses are very heavy are:

\begin{itemize}

\item 3a+3b - box diagrams with one $W$ and one $W^{\prime}$ 
(these are the ones we consider in unitary gauge)
\begin{equation}
(FF)_{ab}=\beta \left[\left(1+\frac{\beta\lambda_{t}\lambda_N}{4}\right)
J_{2}(\lambda_{t},\lambda_N,\beta)-\frac{1+\beta}{4}F_{1}(\lambda_{t},\lambda_N,\beta)\right]
\end{equation}
where the functions $J_{2}$ and $F_{1}$ are defined in \cite{pilaftsis}.

\item 3c+3d+3e+3f - vertex corrections involving the exchange of the 
neutral scalars with tree-level flavor changing couplings.
\begin{eqnarray}
(FF)_{cdef}&=&-\frac{\beta^{2}}{4}\left[
\frac{\lambda_{t}^{2}\log\lambda_{t}}{(1-\lambda_{t})(1-\beta\lambda_{t})} + 
\frac{\lambda_N^{2}\log\lambda_N}{(1-\lambda_N)(1-\beta\lambda_N)} \right .\nonumber \\
&+& \left. \frac{(2\beta\lambda_{t}\lambda_N-\lambda_{t}-\lambda_N)\log\beta}
{(1-\beta)(1-\beta\lambda_{t})(1-\beta\lambda_N)} \right]
+ {\cal O}\left(\frac{1}{M_{H}^{2}}\right)
\end{eqnarray}

\item 1+3g+3h - tree-level exchange of scalars with flavor changing 
couplings as well as self-energy corrections to this.
\begin{equation}
(FF)_{gh}= \frac{\beta}{4}\left(\frac{\log\beta}{1-\beta}+\log\lambda_{\phi}-2\right)+ {\cal O}\left(\frac{1}{M_{H}^{2}}\right)
\end{equation}
where $\lambda_{\phi}= M_{\phi}^{2}/M_{W}^{2}$ and $M_{\phi}$ is the mass of the neutral scalars (assumed degenerate in \cite{pilaftsis}) 
with tree-level flavor changing couplings.
\item all other diagrams considered in Ref.~\cite{pilaftsis} yield contributions to the form factor that vanish as inverse powers of 
$M_{H}$ in the limit of infinite mass for the physical Higgs's that occur.

\end{itemize}
 
The final result in the limit $M_{H} \to \infty$ is then
\begin{eqnarray}
(FF) &\to& \frac{\beta}{4} \left[ 
\log\lambda_{\phi}-2 + 
{\lambda_t\log\lambda_t  \left(\beta\lambda_{t}^{2}-\beta\lambda_{t}-\lambda_{t}+4\right)    \over
(\lambda_N-\lambda_t)(1-\lambda_t)(1-\beta\lambda_t)}\right.
\nonumber \\
&+&\left. 
{\lambda_N\log\lambda_N \left(\beta\lambda_N^{2}-\beta\lambda_N-\lambda_N+4\right)   \over
(\lambda_t-\lambda_N)(1-\lambda_N)(1-\beta\lambda_N)} 
+ 
{3\beta\log\beta \over
(1-\beta)(1-\beta\lambda_t)(1-\beta\lambda_N  )}\right]
\end{eqnarray}

This is to be compared to our result for the same form-factor 
(which appears in Eq.~\ref{fullff}) as obtained 
from Eq.~\ref{tensorfflr}:  $(FF)_{us} \sim (E^{S}_{LR}(\lambda_{t},
\lambda_N,\beta) + 4 E^{T}_{LR}(\lambda_{t},
\lambda_N,\beta))$ and normalized in the same way as the result of 
Ref.~\cite{pilaftsis}:
\begin{eqnarray}
(FF) &=& \frac{\beta}{4} \left[ 
\frac{1}{\hat\epsilon} +\log\left(\frac{\mu^{2}}{M_{W}^{2}}\right)+1 + 
{\lambda_t\log\lambda_t  \left(\beta\lambda_{t}^{2}-\beta\lambda_{t}-\lambda_{t}+4\right)    \over
(\lambda_N-\lambda_t)(1-\lambda_t)(1-\beta\lambda_t)}\right.
\nonumber \\
&+&\left. 
{\lambda_N\log\lambda_N \left(\beta\lambda_N^{2}-\beta\lambda_N-\lambda_N+4\right)   \over
(\lambda_t-\lambda_N)(1-\lambda_N)(1-\beta\lambda_N)} 
+ 
{3\beta\log\beta \over
(1-\beta)(1-\beta\lambda_t)(1-\beta\lambda_N  )}\right]
\end{eqnarray}
Comparing these two results after using our prescription shows that 
we reproduce the complete calculation  in the limit $M_{H}\to \infty$ 
up to the numerical factor $3$ for the choice $\mu=M_{H}$. This 
number depends on the renormalization scheme used and is of the 
same order as the logarithm $\log(\mu^{2}/M_{W}^{2})\sim 5$ for the scale $\mu\sim 1$~TeV we choose.

\end{document}